\documentclass{elsart}
\usepackage{epsfig}

\usepackage{amssymb}

\begin{document}

\begin{frontmatter}

\title{$O(m_d-m_u)$ Effects in CP-even and
CP-odd $K\rightarrow\pi\pi$ Decays}
\author{K. Maltman}
\address{York Univ., Toronto, Canada; CSSM, Univ.
of Adelaide, Adelaide, Australia\thanksref{email1}}
\thanks[email1]{E-mail: maltman@fewbody.phys.yorku.ca}
\author{C.E. Wolfe}
\address{Nuclear Theory Center, Indiana Univ., Bloomington, 
USA\thanksref{email2}}
\thanks[email2]{E-mail: wolfe@niobe.iucf.indiana.edu}

\begin{abstract}
Strong isospin-breaking (IB) effects in CP-even and CP-odd
$K\rightarrow\pi\pi$ decays are computed to next-to-leading
order (NLO) in the chiral expansion.  The impact of these
corrections on the magnitude of the $\Delta I=1/2$ Rule
and on the size of the IB correction, $\Omega_{IB}$,
to the gluonic penguin contribution to $\epsilon^\prime /\epsilon$
are discussed.
\end{abstract}
\end{frontmatter}

In the presence of IB, the standard isospin
decomposition of the $K^+\rightarrow\pi^+\pi^0$, 
$K^0\rightarrow\pi^+\pi^-,\pi^0\pi^0$ decay amplitudes, $A_{+0}$,
$A_{+-}$ and $A_{00}$, becomes~\cite{cdg}
\begin{eqnarray}
\label{isodecomp}
A_{00} &=&  \sqrt{1/3} A_0 
{\rm e}^{i\Phi_0}-[{\sqrt{2/3}}] A_2 {\rm e}^
{i\Phi_2}, \nonumber \\
A_{+-} &=& \sqrt{1/ 3} A_0 
{\rm e}^{i\Phi_0}+[{1/\sqrt{6}}] A_2 {\rm e}^
{i\Phi_2}, \nonumber \\
A_{+0} &=& [{\sqrt{3}/ 2}] A_2^\prime {\rm e}^{i\Phi_2^\prime}.
\end{eqnarray}
In the absence of
the $I=2$ component of electromagnetism (EM),
the $\Phi_I$ are the $\pi\pi$ phases.
In general, $\vert A_2^\prime\vert \not= \vert A_2\vert$ 
due to EM- and strong-IB-induced
$\Delta I=5/2$ contributions.  $A_0$, $A_2$ can be chosen real
in the absence of CP violation.

Since $\vert A_0\vert\sim 20 \vert A_2\vert$,
IB ``leakage'' of the large
octet amplitude into the 
$\Delta I=3/2$ amplitude can be numerically significant.
EM leakage contributions have been
computed to NLO in Ref.~\cite{cdg}; 
we compute the NLO strong octet IB contributions.
These enter
Standard Model predictions of $\epsilon^\prime /\epsilon$
where the strong cancellation between 
gluonic penguin ($O_6$) and electroweak penguin ($O_8$)
contributions is sensitive
to the degree of strong-IB-induced suppression of the $O_6$ 
contribution~\cite{burasrev}.

At leading chiral order (LO), the computation of the
octet leakage contribution is unambiguous;
the magnitude of the LO weak $27$-plet low-energy constant
(LEC) is {\it decreased} by $\Omega_{IB}=13\%$.  
The corresponding 
$O_6$ suppression in $\epsilon^\prime /\epsilon$
is $1-\Omega_{IB}$.  Recent analyses of $\epsilon^\prime /\epsilon$
employ $\Omega_{IB}=0.25\pm 0.08$, the difference from the LO
value reflecting estimates of the effect of 
$\eta^\prime$ mixing.  This effect is NLO in the chiral
expansion, but does not exhaust NLO contributions.
A full NLO calculation can be performed using
Chiral Perturbation Theory (ChPT).
The importance of such a {\it complete}
NLO determination can be
seen from the recent discussion of NLO
$\pi -\eta$ mixing effects ~\cite{emnp99}:
the $\eta^\prime$
contribution (associated with the strong 
LEC $L_7^r$) turns out to be
almost completely cancelled by a contribution proportional
to $L_8^r$~\cite{emnp99}.
To compute the NLO IB leakage contributions
one evaluates the tree and
one-loop graphs of Ref.~\cite{cwkmk2pi}. 
NLO tree contributions are either proportional to
the product of the LO weak octet LEC $c^\pm$ and a single NLO strong LEC
or proportional to one of the NLO
weak LEC's. All loop graphs
involve one vertex from the LO octet effective weak Lagrangian,
$c^{\pm}Tr\left[ \lambda^{\pm}\partial_\mu U^\dagger
\partial^\mu U\right]$,
where the superscripts $\pm$ label the CP even and odd cases,
respectively, $\lambda^+=\lambda_6$, $\lambda^-=\lambda_7$,
and $U=exp\left( i{\lambda}\cdot{\pi}\right)$, is the
usual matrix variable.
The (scale-dependent) ratio of the sum of the loop contributions to the
LO octet contribution for a given amplitude is thus completely
fixed; the main uncertainty lies in
a lack of knowledge of the NLO weak
LEC's, for which we are forced to use models
(see Refs.~\cite{cwkmk2pi,cwkmee} for further discussion).

\begin{table}
\caption{Strong octet and EM IB leakage contributions in
units of $10^{-6}\ {\rm MeV}$. The IC and
LO IB fits
yield $A_2 = A_2^\prime = -2.1\times 10^{-5}\ {\rm MeV}$
and $-2.4\times 10^{-5}$ MeV, respectively. }\label{table1}
\begin{tabular}{lcc}
\hline
Source\ \ \qquad\qquad&$\delta^{(s)}A_2$&
$\delta^{(s)}A_2^\prime$ \\ 
\hline
$(8)\qquad$&$(-1.56\pm 0.63) +(0.42\pm 0.05){\rm i}$&
$(-1.56\pm 0.63)+(0.42\pm 0.05){\rm i}$ \\
$(EM)$\qquad&$(-1.27\pm 0.40)
-(1.28\pm 0.02){\rm i}$&$(0.70\pm 0.73)-(0.07\pm 0.04){\rm i}$ \\
\hline
\end{tabular}
\end{table}

The contributions to $A_2$ and $A_2^\prime$ associated with
EM~\cite{cdg} and octet IB~\cite{cwkmk2pi} leakage are given
in Table 1.  
The errors reflect uncertainties in the estimates of the
NLO LEC's.  Denoting the ratio of LO
$27$-plet to octet weak LEC's obtained neglecting,
or including, IB by $r_{IC}$, or $r_{IB}$, respectively,
we find 
$R_{IB}\equiv r_{IB}/r_{IC}=0.963\pm 0.029\pm 0.010\pm 0.034$.
The errors reflect uncertainties in
the weak NLO LEC combinations, the input value of
$B_0(m_d-m_u)$, and the EM contributions, respectively.  
The deviation from $1$
is significantly smaller than at LO (where $R_{IB}=0.870$).
The $\Delta I=5/2$ contribution (dominantly EM in
character~\cite{cwkmk2pi}), leads to
${\vert A_2\vert}/{\vert A_2^\prime \vert}=1.094\pm 0.039\not= 1$,
and significantly exacerbates the phase discrepancy problem
for the neutral $K$ decays~\cite{cwkmk2pi}.

For the CP-odd case, 
$\Omega_{IB}=\omega \, {Im\, \delta A_2}/\, {Im\, A_0}$
($\omega= Re\, A_0/Re\, A_2\simeq 22.2$;
$\delta A_2$ is the octet leakage contribution).
At LO, $\Omega_{IB}=0.13\equiv [\Omega_{IB}]_{LO}$.
At NLO
$\Omega_{IB}=[\Omega_{IB}]_{LO}\left[
1+{\frac{{\rm Im}\, \delta A_2^{(NLO;ND)}}
{{\rm Im}\, \delta A_2^{(LO)}}}-{\frac{{\rm Im}\, A_0^{(NLO;ND)}}
{{\rm Im}\, A_0^{(LO)}}}\right]\equiv
[\Omega_{IB}]_{LO}\left[ 1+R_2-R_0\right]$.
The superscript $(NLO;ND)$ indicates the sum of non-dispersion
NLO contributions (involving NLO weak and strong LEC's
and the non-dispersive parts of loop graphs).  
Neither the NLO $I=0$ IC
nor NLO $I=2$ IB leakage CP-odd LEC combinations are known.
The NLO dispersive contributions create phases consistent
with Watson's theorem.
Although the positive $I=0$ phases correspond to
attractive FSI, 
NLO weak LEC corrections may, nonetheless,
make ${Im}\, A_0$ smaller at NLO than the LO
(see comments on Ref.~\cite{pichpallente} 
in Ref.~\cite{burasetal}
for a related discussion).
If, however, NLO effects {\it do} enhance $Im\, A_0$ 
(decreasing the level of $O_6$-$O_8$ cancellation and
increasing $\epsilon^\prime /\epsilon$) 
$\Omega_{IB}$ will be simultaneously suppressed,
further increasing $\epsilon^\prime /\epsilon$.
The known NLO contributions (loops and strong
LEC terms) give contributions $-0.24 (-0.31)$ to $R_2$
and $-0.02 (+0.42)$ to $R_0$,
at scale $\mu =m_\eta (m_\rho )$.  
Using the weak deformation model to estimate the weak NLO LEC's,
$1+R_2-R_0=0.27$ while, for the chiral quark model, it
lies between $0.62$ and $1.42$.
Averaging these
results, and taking their spread as a
{\it minimal} indication of theoretical error, we thus
obtain, at NLO,
$\Omega_{IB}=0.11\pm 0.08$, a
much smaller value than conventionally employed,
though with comparable errors.  It is also significantly smaller
than the partial NLO estimate of Ref.~\cite{emnp99},
$0.16\pm 0.03$, based on the strong
LEC contributions only.
The central value above, combined
with conventional central values for the $B$-factors,
leads to a $\sim 50\%$ increase in the 
predicted value for $\epsilon^\prime /\epsilon$.  
To be conservative, we would propose using this
lower central value 
with an even larger error estimate, in all future calculations
of Standard Model values for $\epsilon^\prime /\epsilon$. 

\end{document}